\documentclass{appolx}
\usepackage{epsfig}
\begin{document}
\title{Reentrant transitions of Ising-Heisenberg ferromagnet on a triangular lattice
with diamond-like decorations\thanks{Presented at CSMAG'07 Conference, Ko\v{s}ice, 9-12 July 2007}}
\author{M.~JA\v{S}\v{C}UR, J.~STRE\v{C}KA and L.~\v{C}ANOV\'A
\address{Department of Theoretical Physics and Astrophysics, Faculty of Science, \\ 
P. J. \v{S}af\'{a}rik University, Park Angelinum 9, 040 01 Ko\v{s}ice, Slovak Republic}}
\maketitle
\begin{abstract}
The mixed spin-1/2 and spin-1 Ising-Heisenberg ferromagnet on the decorated triangular lattice 
consisting of inter-connected diamonds is investigated within the framework of an exact decoration-iteration mapping transformation. It is shown that the diamond-like decoration 
by a couple of the Heisenberg spins gives rise to a diverse critical behaviour including 
reentrant phase transitions with two consecutive critical points.
\newline
\end{abstract}
\PACS{05.50.+q, 68.35.Rh}

\section{Introduction}
The quantum Heisenberg model represents a long-standing theoretical challenge in the condensed 
matter physics especially due to insurmountable mathematical complexities associated with 
a non-commutability of spin operators involved in its Hamiltonian. The non-commutability between 
the relevant spin operators raises a zero-point motion (quantum fluctuations) and consequently, low-dimensional antiferromagnets often exhibit a variety of exotic ground states owing to quantum 
phase transitions driven by the quantum fluctuations. By contrast, low-dimensional 
ferromagnets are usually thought of as being less affected by the quantum fluctuations. 

Several low-dimensional quantum Heisenberg models can exactly be solved merely on behalf of 
their special geometries \cite{lieb62,maju69}. Simple geometries, which are solely composed of inter-connecting diamonds, turned out to be a very useful testing ground for elucidating quantum properties of the low-dimensional spin systems \cite{long90}. In our previous work, we have proposed 
the Ising-Heisenberg model on decorated planar lattices consisting of inter-connected diamonds 
in order to clarify a mutual interplay between the long-range magnetic order and quantum fluctuations \cite{stre02}. The rigorous solution revealed a surprisingly rich ground-state phase diagram 
even for the ferromagnetic model, which consists of the classical ferromagnetic phase (CFP), 
the quantum ferromagnetic phase (QFP) and the disordered phase (DP). It is worthwhile to remark, moreover, that DP originates from a peculiar geometric spin frustration related to the 
competition between the easy-axis and easy-plane interactions. Therefore, the purpose of 
the present work is to shed light on how this competition influences (dis)appearance of 
spontaneous ordering of two long-range ordered phases CFP and QFP.

\section{Model and exact map}
First, let us briefly describe the hybrid Ising-Heisen\-berg model on the decorated 
triangular lattice constituted by inter-connected diamonds as shown in Fig.~1. 
\begin{figure}[t]
     \begin{center}
       \includegraphics[width = 0.67\textwidth]{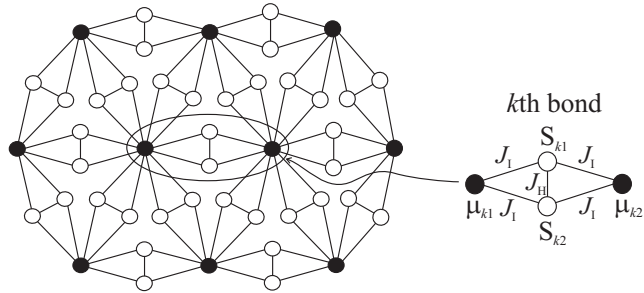}
       \caption{\small The cross-section of the decorated triangular lattice consisting of  
       inter-connected diamonds. The full and empty circles illustrate lattice positions 
       of the Ising and Heisenberg spins, respectively. The ellipse allocates an elementary 
       cell (spin cluster) of the decorated lattice described through 
       the bond Hamiltonian (\ref{1}).}
       \label{fig1}
     \end{center}
\end{figure}
It is quite obvious that the considered lattice is formed by two different spin sites 
schematically depicted in Fig.~1 as full and empty circles, respectively. The former 
spin sites are occupied by the spin-1/2 Ising atoms described by spin operators 
$\hat{\mu}_{k \alpha}^{z}$ ($\alpha =1,2$), while the latter ones are occupied 
by the spin-1 Heisenberg atoms described by spin operators $\hat{S}_{k \alpha}^{\gamma}$ 
($\gamma = x,y,z$). The total Hamiltonian can be for convenience defined as a sum of bond 
Hamiltonians, $\hat{{\cal H}} = \sum_{k} \hat{{\cal H}}_k$, where each bond Hamiltonian
\begin{eqnarray}
\hat{{\cal H}}_k = \! \! \! &-& \! \! \!  J_{\rm H} [\Delta (\hat{S}_{k1}^x \hat{S}_{k2}^x + \hat{S}_{k1}^y \hat{S}_{k2}^y) + \hat{S}_{k1}^z \hat{S}_{k2}^z] - D [(\hat{S}_{k1}^z)^2 + (\hat{S}_{k2}^z)^2] \nonumber \\ \! \! \! &-& \! \! \!
J_{\rm I} (\hat{S}_{k1}^z + \hat{S}_{k2}^z) (\hat{\mu}_{k1}^z + \hat{\mu}_{k2}^z)
\label{1}	
\end{eqnarray}
involves all the interactions terms of one diamond-like spin cluster residing on the $k$th bond of 
the triangular lattice. Above, the parameter $J_{\rm H}(\Delta)$ represents the anisotropic 
XXZ Heisenberg interaction between the nearest-neighbouring spin-1 atoms, $\Delta$ is a 
spatial anisotropy in this exchange interaction, whereas the parameter $D$ measures a strength 
of the uniaxial single-ion anisotropy acting on the spin-1 atoms. Finally, the parameter 
$J_{\rm I}$ denotes the Ising-like exchange interaction between the nearest-neighbouring 
spin-1/2 and spin-1 atoms, respectively. 

The aforedescribed Ising-Heisen\-berg model can accurately be treated by adopting the well-known 
procedure based on the exact decoration-iteration mapping transformation \cite{stre02}.
As a result of this mapping, the partition function ${\cal Z}$ of the mixed spin-1/2 and spin-1 Ising-Heisenberg model on the diamond-like decorated triangular lattice can be expressed in terms of the partition function ${\cal Z}_0$ of the spin-1/2 Ising model on the simple triangular lattice 
with some effective nearest-neighbour interaction $R$
\begin{eqnarray}
{\cal Z} (T, J_{\rm H}, \Delta, D, J_{\rm I}) = A^{3N} {\cal Z}_{0} (T, R). 
\label{2}	
\end{eqnarray}
Note that both mapping parameters $A$ and $R$ are unambiguously determined by a self-consistency condition of the mapping transformation and their explicit form is given by Eqs.~4--7 of Ref. \cite{stre02}. It directly follows from the relation (\ref{2}) that the hybrid Ising-Heisenberg 
model becomes critical if and only if its corresponding spin-1/2 Ising model on the triangular 
lattice becomes critical, as well. Accordingly, the exact critical temperature $T_{\rm C}$ of the hybrid Ising-Heisenberg model must obey the condition $k_{\rm B} T_{\rm C} = R/\ln 3$ 
($k_{\rm B}$ is Boltzmann's constant). 

\section{Results and discussions}
\begin{figure}[t]
     \begin{center}
       \includegraphics[width = 0.52\textwidth]{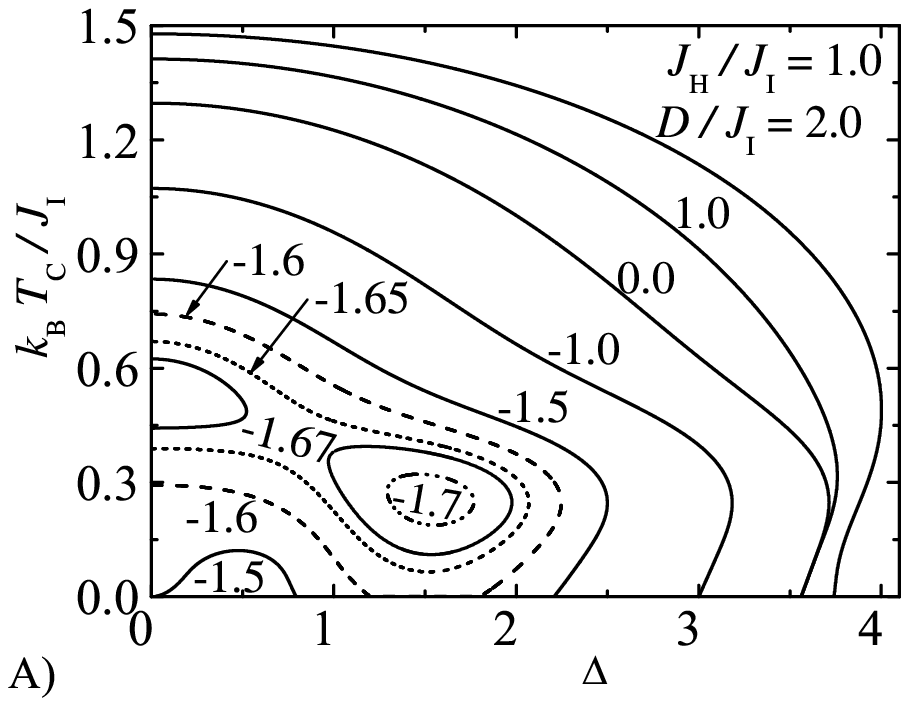}
       \hspace*{-8mm}
       \includegraphics[width = 0.52\textwidth]{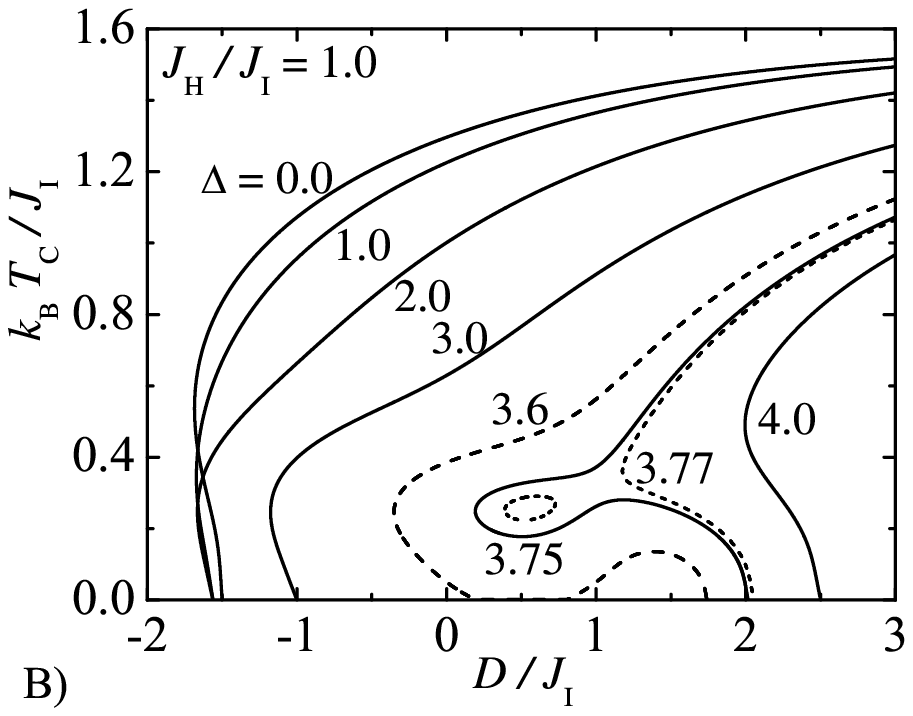}
       \caption{\small A) The critical temperature as a function of the exchange anisotropy 
       $\Delta$ for several values of the single-ion anisotropy $D/J_{\rm I}$ and 
       $J_{\rm H}/J_{\rm I} = 1.0$. B) The critical temperature as a function of the single-ion 
       anisotropy $D/J_{\rm I}$ for several values of the exchange anisotropy $\Delta$ and 
       $J_{\rm H}/J_{\rm I} = 1.0$.}
       \label{fig2}
     \end{center}
\end{figure}

The dependence of reduced critical temperature on the anisotropy parameters $\Delta$ and $D/J_{\rm I}$ is shown in Fig.~\ref{fig2} by keeping the ratio $J_{\rm H}/J_{\rm I} = 1.0$ constant. 
As it can be clearly seen, the model under investigation displays a rather complex critical
behaviour that depends very sensitively on a strength of both anisotropy parameters.
Within the finite-temperature phase diagram one actually finds several regions where the 
system exhibits reentrant phase transitions with two consecutive critical temperatures, which 
might be even relatively well separated. Generally, the reentrance presumably occurs either if the 
single-ion anisotropy is of an easy-axis type ($D>0$) and the exchange anisotropy is of an easy-plane
type ($\Delta > 1$) or if the reverse is the case. This indicates that the reentrance is closely related to a competition between the easy-plane and easy-axis interactions, while the latter one 
is also supported by the Ising interaction $J_{\rm I}$. Besides, it is worthwhile to remark
that the most interesting part of the finite-temperature phase diagram includes closed loops of 
phase transition lines, which separate the long-range ordered phase (either CFP or QFP) from 
the disordered one. Notice that the results displayed in Fig.~\ref{fig2} have been convincingly evidenced also by a detailed analysis of the spontaneous magnetization, which represents the 
order parameter for both ferromagnetic phases CFP and QFP.    

In conclusion, it should be mentioned that the critical behaviour of the Ising-Heisenberg models 
on different diamond-like decorated planar lattices basically depends also on a lattice topology 
even though all lattices have the same ground-state phase diagram regardless of their coordination number \cite{stre02}. In this respect, more systematic study that would clarify this issue 
in detail is left for our future work.

\begin{center}
{\bf Acknowledgments}
\end{center}
This work was supported by the Slovak Research and Development Agency 
under the contract No. LPP-0107-06 and the grant VVGS PF 02/2007/F.

\end{document}